\documentclass[sn-mathphys,Numbered,iicol,colorlinks=true,linkcolor=blue,citecolor=blue, urlcolor=blue]{sn-jnl}
\pdfoutput=1

\usepackage{graphicx}%
\usepackage{multirow}%
\usepackage{amsmath,amssymb,amsfonts}%
\usepackage{amsthm}%
\usepackage{mathrsfs}%
\usepackage[title]{appendix}%
\usepackage{xcolor}%
\usepackage{textcomp}%
\usepackage{manyfoot}%
\usepackage{booktabs}%
\usepackage{algorithm}%
\usepackage{algorithmicx}%
\usepackage{algpseudocode}%
\usepackage{listings}%

\usepackage{hyperref}   
\usepackage{tikz,calc}
\usetikzlibrary{shapes,patterns,arrows,positioning,automata}
\usepackage{pgfplots}
\usepackage{units}
\usepackage{ulem}
\usepackage[T1]{fontenc}
\usepackage{enumitem}  
\usepackage{xspace}

\def\Eq#1{Eq.~(\ref{#1})}

\def\Fig#1{Fig.~\ref{#1}}

\usepackage{bm}
\def\p{\mathbf{p}}
\def\phat{\hat{\mathbf{p}}}

\def\x{\mathbf{x}}

\def\k{\mathbf{k}}

\def\pp{\mathbf{p'}}

\def\kp{\mathbf{k'}}

\newcommand{\kor}[1]{{\color{blue} #1}}

\usepackage{dcolumn}

\renewcommand{\d}{\text{d}}

\newcommand{\eVnospace}{\text{e\kern-0.15ex V}\xspace}
\newcommand{\eV}{\text{ e\kern-0.15ex V}\xspace}

\newcommand{\GeV}{\text{ G\eVnospace}\xspace}

\begin{document}


\title[Thermalization and isotropization in the AMY parton cascade \textsc{Alpaca}]{Thermalization and isotropization in the AMY parton cascade \textsc{Alpaca}}



\author[1]{\fnm{Robin} \sur{Törnkvist}}\email{robin.tornkvist@hep.lu.se}

\author[1]{\fnm{Korinna} \sur{Zapp}}\email{korinna.zapp@hep.lu.se}


\affil[1]{\orgdiv{Department of Physics
}, \orgname{Lund University}, \orgaddress{\street{Box 118, SE-221 00 Lund}, \country{Sweden}}}


\abstract{We look at thermalization and isotropization processes in the newly introduced AMY QCD kinetic theory parton cascade \textsc{Alpaca}. For thermalization, we consider the case of overoccupied initial conditions, and study the time evolution of the distribution as it relaxes to thermal equilibrium. We find that the system thermalizes as expected compared to known analytical results. For anisotropic systems, we take a first look at the qualitative behaviour of isotropization for Color Glass Condensate-like initial conditions in a homogeneous box with periodic boundary conditions.}

\maketitle


\section{Introduction}
\label{section:introduction}

In the pre-equilibrium stage of heavy ion collisions, the equilibration processes which relax the system to a point where viscous hydrodynamics is applicable are a central topic of study. These processes can be understood through the dynamics of both strong coupling \cite{Chesler:2009cy,Heller:2011ju,Chesler:2010bi,Casalderrey-Solana:2013aba} as well as weak coupling \cite{Baier:2000sb,Baier:2002bt,Berges:2013eia}. For the case of weak coupling, the QCD effective kinetic theory introduced by P.~Arnold, G.~Moore and L.~Yaffe (AMY) in \cite{Arnold:2002zm} emerges as an excellent candidate to model the evolution, as it is valid both in- and out-of-equilibrium. Furthermore, it has been shown to lead to rapid equilibration~\cite{Kurkela:2015qoa}.

The AMY effective kinetic theory provides a leading-order (in 't Hooft coupling coupling $\lambda$) description of the evolution of the phase space densities $f$, for temperatures $T$ where the coupling $g(T)$ is small. The relevant processes at leading order are elastic scattering and (quasi-)collinear merging/splitting, where the latter includes a QCD version of the LPM effect. It has been studied extensively for small collision systems, see e.g.~\cite{Kurkela:2018qeb,Kurkela:2019kip,Kurkela:2021ctp,Ambrus:2021fej}.

In \cite{Kurkela:2022qhn} we introduced \textsc{Alpaca} (AMY Lorentz invariant PArton CAscade), a parton cascade which evolves discrete parton ensembles according the the dynamics of AMY. In the same paper we also validate the system in thermal equilibrium. We now extend the validation of the framework by looking at thermalization and isotropization in \textsc{Alpaca}. Many studies of equilibration in AMY kinetic theory have been done, e.g. \cite{Kurkela:2014tea, Kurkela:2015qoa, Fu:2021jhl, Kurkela:2018oqw}. In this article, we will look at thermalization of overoccupied initial conditions, and compare our results of the evolution of $f$ as well as the time of equilibration to previously known results. We will also take a first look at isotropization in \textsc{Alpaca}, using Color Glass Condensate-like initial conditions in a homogeneous box with periodic boundary conditions.

\section{AMY Effective Kinetic Theory}
\label{section:amy}

In \cite{Arnold:2002zm} P.~Arnold, G.~Moore and L.~Yaffe (AMY) introduce an effective kinetic theory \kor{of QCD} which describes the evolution of (spin and color averaged) distribution functions for gluons and quarks through the Boltzmann equations
\begin{equation}
    \label{eq:AMY_Boltzmann}
    \left(\partial_t + \frac{\p}{p_0} \cdot \nabla_\mathbf{x} \right) f_s(\x,\p,t) = - C_s^{2\leftrightarrow 2}[f] - C_s^{``1\leftrightarrow 2"}[f]
\end{equation}
for a particle of species $s$. The evolution is described to leading order in the coupling $\lambda = g^2 N_c = 4\pi\alpha_s N_c$ for sufficiently high temperatures $T$ where the coupling $g(T)$ is small. This is valid under the assumption that the two energy scales set by $T$ (hard) and $g(T)$ (soft) are well separated, both parametrically and quantitatively. The collision kernels that appear in \Eq{eq:AMY_Boltzmann} encode all effects which, to leading order, contribute to the evolution. These effect are elastic scattering, described by $C_s^{2\leftrightarrow 2}$, and (quasi-)collinear splitting and merging described by $C_s^{``1\leftrightarrow 2"}$.

The elastic scattering kernel corresponding to a particle of species $a$ is given by
\begin{align}
    & C_a^{2\leftrightarrow2}[f]  =  \frac{1}{4|\p|} \sum_{bcd} \int_{\k\pp\kp} \delta^{(4)}(P+K-P'-K') \nonumber\\ 
    &  \times \nu_b |\mathcal{M}^{ab}_{cd} |^2 (2\pi)^4\big\{ f_a(\p)f_b(\k)[1\pm f_c(\pp)][1\pm f_d(\kp)] \nonumber \\ 
    & - f_c(\pp)f_d(\kp)[1\pm f_a(\p)][1\pm f_b(\k)] \big\} \,, 
    \label{eq:AMY_C_elastic}
\end{align}
where $\nu_b$ is the number of spin and colors states of $b$, and the shorthand notation 
\begin{equation}
    \int_{\p} \dots = \int\frac{\d^3 p}{2|\p| (2\pi)^3} \dots
\end{equation}
is used. The terms $|\mathcal{M}^{ab}_{cd}|^2$ are the squared  $2\leftrightarrow2$ matrix elements summed over the final state and averaged over the initial state\footnote{There is a difference in factors of $nu_s$ compared to \cite{Arnold:2002zm} since they there take the squared matrix elements to be summed over both initial and final state.}. The matrix elements are divergent in the $t$ and $u$ channels, which is regulated by leading order screening effects through the effective mass, which for gluons is defined as
\begin{equation}
    \label{eq:AMY_mg2}
    m_g^2 = \sum_{s} 2\nu_s \frac{g^2 C_s}{d_A} \int_V \frac{d^3\x}{V} \int \frac{d^3\p}{2|\p|(2\pi)^3}f_s(\p,\x)
\end{equation}
where $d_A=8$ is the dimension of the adjoint representation of the gluons, and $C_A=3$, $C_F=4/3$ are the quadratic Casimirs.  

The collinear splitting and merging is encoded in 
\begin{align}
    & C_a^{``1\leftrightarrow 2"}[f] =  \frac{(2\pi)^3}{2|\p|^2\nu_a}\sum_{b,c}\int_0^\infty \d p'\,\d k'\delta(p-p'-k') \nonumber \\ 
    & \ \times \gamma^a_{bc}(\p;p'\phat,k'\phat)\Big\{f_a(\p)[1\pm f_b(p'\phat)][1\pm f_c(k'\phat)] \nonumber \\ 
    &  \qquad \qquad \qquad \qquad - f_b(p'\phat)f_c(k'\phat)[1\pm f_a(\p)] \Big\} \nonumber \\
    & + \frac{(2\pi)^3}{|\p|^2\nu_a}\sum_{b,c}\int_0^\infty \d k\,\d p'\delta(p+k-p') \nonumber \\ 
    &  \times \gamma^a_{bc}(p'\phat;\p,k\phat) \Big\{f_a(\p)f_b(k\phat)[1\pm f_c(p'\phat)] \nonumber \\
    & \qquad \qquad \qquad \quad - f_c(p'\phat)[1\pm f_a(\p)][1\pm f_b(k\phat)] \Big\},     \label{eq:AMY_C_inelastic_collinear}
\end{align}
where $\gamma^a_{bc}$ is the differential splitting/merging rate. Collinear splitting/merging is only allowed by involving a soft interaction with the medium, and the process is not instantaneous and multiple soft scatterings can occur during the formation time of the process. This turns the splitting/merging into a $N+1 \leftrightarrow N+2$ process, but the additional $N$ soft scatterings are not resolved individually. They rather act coherently, which gives rise to a QCD analogue of the Landau-Pomeranchuk-Migdal (LPM) effect. This effect is accounted for in the splitting/merging rate $\gamma$. The splitting/merging rate is also a function of  the effective temperature $T_*$, defined as 
\begin{align}
    \label{eq:Tstar}
    T_* & = \frac{\frac{1}{2}g^2 \sum_s\frac{\nu_s C_s}{d_A}\frac{1}{V}\int d^3\x \int \frac{d^3\p}{(2\pi)^3}f_s(p)[1+f_s(p)]}{g^2\sum_s\frac{ \nu_s C_s}{d_A}\frac{1}{V}\int d^3\x \int\frac{d^3\p}{(2\pi)^3}f_s(p)/p} \nonumber  \\ 
    & =  \frac{T_*'}{m_g^2}.
\end{align}
This effective temperature corresponds to the temperature of a thermal system  which would have the same scattering rate as the system under consideration.

\section{\textsc{Alpaca}}
\label{section:alpaca}

In \cite{Kurkela:2022qhn} we introduce \textsc{Alpaca} (AMY Lorentz invariant PArton CAscade) which is a parton cascade that indirectly solves \Eq{eq:AMY_Boltzmann} by evolving a parton ensemble according to the AMY collision kernels. The parton cascade is constructed as a module to the multi-purpose event generator \textsc{Sherpa}\cite{Sherpa:2019gpd}. The evolution is implemented in a Lorentz invariant way by regarding the particles in $8N$ dimensional space\footnote{This to avoid the no-interaction theorem \cite{Currie:1963rw} which states that particles moving in a $6N$ dimensional phase space cannot interact if Lorentz invariance is required.}. We let our system evolve in a Lorentz scalar $\tau$~\cite{Peter:1994yq,Borchers:2000wf}, ordering all possible scatterings and mergings of particles $i$ and $j$ based on the corresponding $\tau_{ij}$ where they have their closest approach. The closest approach is defined through the Lorentz invariant measure
\begin{equation}
    \label{eq:ALPACA_LI_distance}
 d_{ij}^2 = - \left( x^2 - \frac{(x_\mu p^\mu)^2}{p^2} \right) \,.
\end{equation}
Here, $x = x_i-x_j$ is the relative four-distance between the particles, and $p = p_i+p_j$ is the total four-momentum of the pair. At each closest approach a total cross section $\sigma_{ij}$ for the scattering and merging is calculated, and the process will happen if $d_{ij}^2<\sqrt{\sigma_{ij}/\pi}$. The cross sections depend on the dynamic quantities $m_{g/q}^2$, $T_*$ and $f(\x,\p)$ which are extracted locally from the ensemble. The splitting of particles does not involve incoming particle pairs and so it cannot follow the same scheme, it is instead implemented using the Sudakov Veto Algorithm. A detailed discussion of the implementation of all processes can be found in \cite{Kurkela:2022qhn}.

The framework has been extensively validated in the case of thermal equilibrium of infinite spatial size. This is done through initialising the system in a box with periodic boundary conditions, which introduces certain technical challenges since the Lorentz invariant distance depends on the momentum of the particle pair. The same issues appear when looking at thermalization and isotropization of infinitely sized systems, which is the main focus of this article. These problems, and how to solve them, are discussed in detail in \cite{Kurkela:2022qhn}.

\section{Results}
\label{section:results}

We will in this paper extend our validation of \textsc{Alpaca} by looking at two equilibration processes;  thermalization in overoccupied systems and isotropization with anisotropic Color Glass Condensate-like initial conditions. Both these two processes are relevant for modelling the early stages of heavy ion collisions, and so it is important that \textsc{Alpaca} manages to capture the expected behaviour. We begin with the thermalization.

\subsection{Thermalization}
We follow \cite{Kurkela:2014tea, Fu:2021jhl} and take our initial condition as a fixed point solution of the overoccupied distribution given in \cite{AbraaoYork:2014hbk}, which has the form
\begin{equation}
    \label{eq:fOO}
    f_{\mathrm{OO}}(p) = \frac{1}{\left(Q\tilde{t}\right)^{4/7}\lambda}\tilde{f}(\tilde{p}).
\end{equation}
Here
\begin{equation}
    \tilde{f}(\tilde{p}) = \frac{0.22e^{-13.3\tilde{p}} + 2.0e^{-0.92\tilde{p}^2}}{\tilde{p}},
\end{equation}
with the rescaled momentum defined as $\tilde{p} = (p/Q)(Q\tilde{t})^{-1/7}$, and $Q$ is the characteristic energy scale. This distribution is valid for early times as long as the typical momentum scale is small compared to the thermal scale, i.e. when $\langle p \rangle \ll T$ and $\tilde{t}\ll t_{\mathrm{eq}}$, where $t_{\mathrm{eq}}$ is the thermal equilibration time. Hence, we treat $\tilde{t}$ as a constant and pick it such that $\tilde{t}=t_0\ll t_{\mathrm{eq}}$, which gives us a fixed point solution to initialize our system in \textsc{Alpaca} with. The system initially consists only of gluons. For consistence with~\cite{Kurkela:2014tea, Fu:2021jhl} we don't produce quarks during the evolution leaving the question of chemical equilibration to a dedicated study.

To determine when the system has thermalized, we follow \cite{Fu:2021jhl} and define the effective temperatures
\begin{equation}
    T_\alpha = \left[\frac{2\pi^2}{\Gamma(\alpha+3)\zeta(\alpha+3)}\int \frac{d^3\x}{V} \int\frac{d^3\p}{(2\pi)^3}p^\alpha f(p) \right]^{\frac{1}{\alpha+3}},
\end{equation}
where we have the added the spatial integral and corresponding normalization to spatial volume $V$, since we will integrate over a finite box. This allows us to define the kinetic thermalization time $t_{\mathrm{eq}}$ as
\begin{equation}
    \left(\frac{T_0(t_{\mathrm{eq}})}{T_1(t_{\mathrm{eq}})}\right)^{-4} = 0.9.
\end{equation}

Since the system under consideration is homogeneous we will utilize the same setup of a box with periodic boundary conditions in \textsc{Alpaca}, as described in \cite{Kurkela:2022qhn}. We will focus on the parameters $\lambda=1$, $t_0 = 0.016$ $\GeV^{-1}$ and $Q=0.5$ $\GeV$, since it gives us a initial distribution which closely resemble the one shown in \cite{Fu:2021jhl}. We will initialize the system in a box of size $1205$ $\GeV^{-3}$ and set a momentum cutoff at $p_{\mathrm{min}}=0.1$ \GeV. These parameters are used for all the data and figures presented below in this subsection.

\medskip

\begin{figure}[tbp]
    \centering
    \includegraphics[width=1\linewidth]{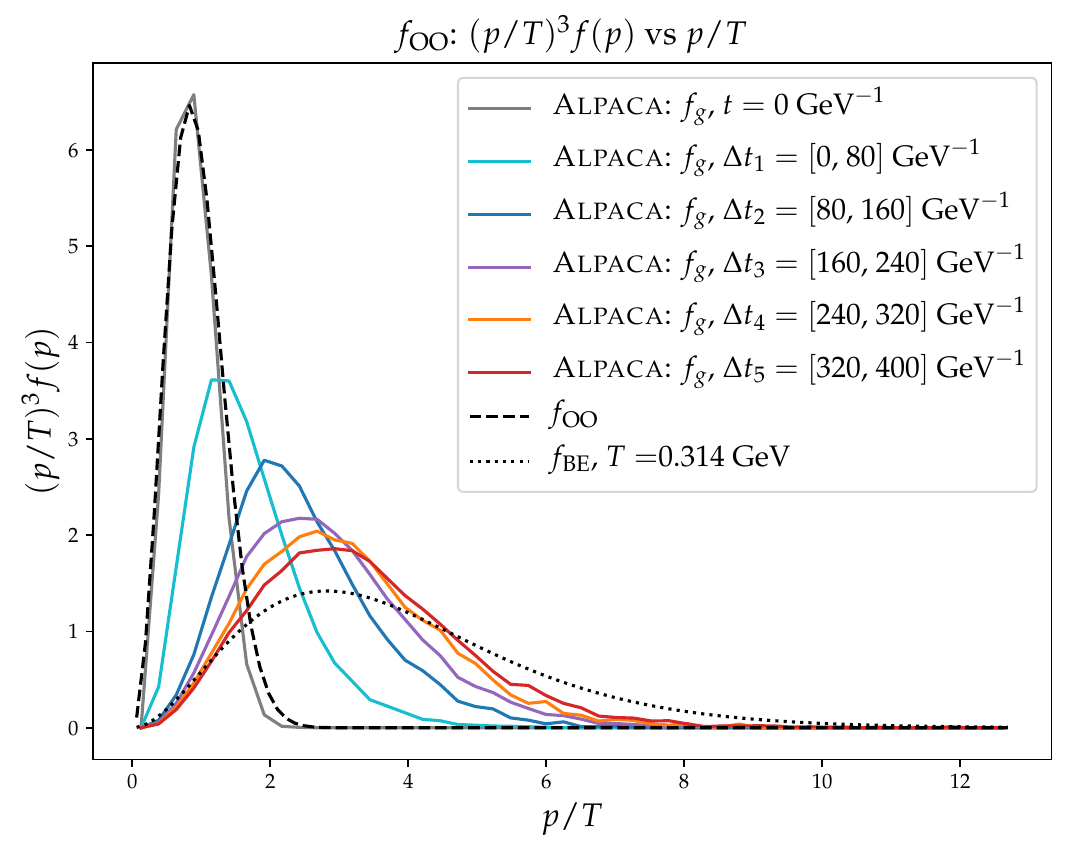}
    \caption{Evolution of the weighted distribution $(p/T)^3f(p)$ as a function of $p/T$. The dashed line indicate the analytic initial condition given in \Eq{eq:fOO}, while the dotted line represent the thermal equilibrium distribution with a temperature corresponding to the effective temperature $T_1$ for the overoccupied distribution. The gray solid line is the weighted distribution sampled in \textsc{Alpaca} at $t=t_0$. The remaining solid lines are the weighted distributions averaged over different time intervals shown in the legend.}
    \label{fig:thermalOO_p3f}
\end{figure}

In \Fig{fig:thermalOO_p3f} the evolution of $(p/T)^3f(p)$ in \textsc{Alpaca} is shown for the overoccupied system. As can be seen, it closely resembles the behaviour shown in \cite{Fu:2021jhl} for an overoccupied system approaching the thermal equilibrium distribution, with a simultaneous decrease in amplitude and shift to harder momentum.

\begin{figure}[tbp]
    \centering
    \includegraphics[width=1\linewidth]{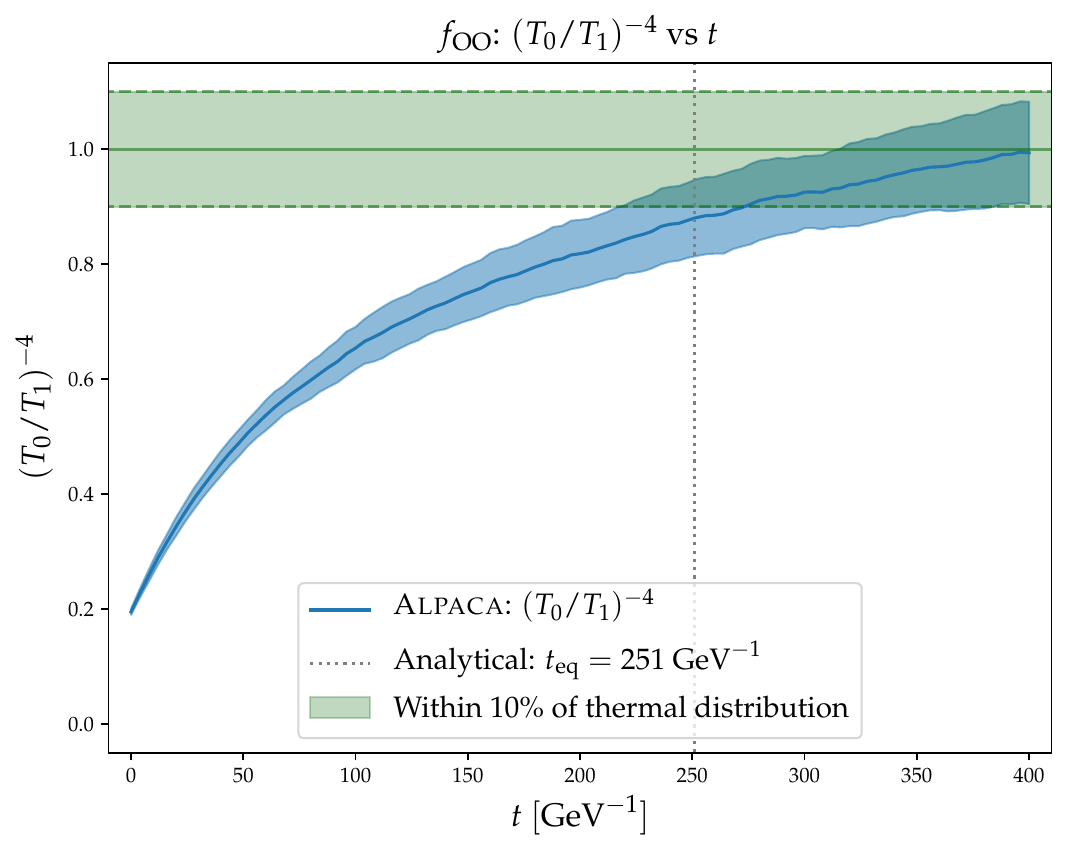}
    \caption{The ratio $(T_0/T_1)^{-4}$ as a function of time $t$, for the overoccupied system in \textsc{Alpaca}. The blue band corresponds to the error between events. The green band corresponds to values within $10\%$ of unity. The dashed line is the equilibration time extracted from \Eq{eq:fOO_teq}. }
    \label{fig:thermalOO_T0T1}
\end{figure}

For a more quantitative comparison to the direct integration of the Boltzmann equation we calculate the ratio of effective temperatures $(T_0/T_1)^{-4}$ used in~\cite{Fu:2021jhl} to define the thermalisation time. The ration is shown in \Fig{fig:thermalOO_T0T1} as a function of time. The equilibration time $t_{\mathrm{eq}}$ is defined as the time where $(T_0/T_1)^{-4} = 0.9$. The equilibration time found in \textsc{Alpaca} is consistent with the value of $\unit[251]{GeV^{-1}}$ found from 
\begin{equation}
    \label{eq:fOO_teq}
    t_{\mathrm{eq}} = \frac{1}{T\lambda^2}\frac{76}{1-0.19\log\lambda}
\end{equation}
in~\cite{Fu:2021jhl} and shown as the vertical line in \Fig{fig:thermalOO_T0T1}.

\begin{figure}[tbp]
    \centering
    \includegraphics[width=1\linewidth]{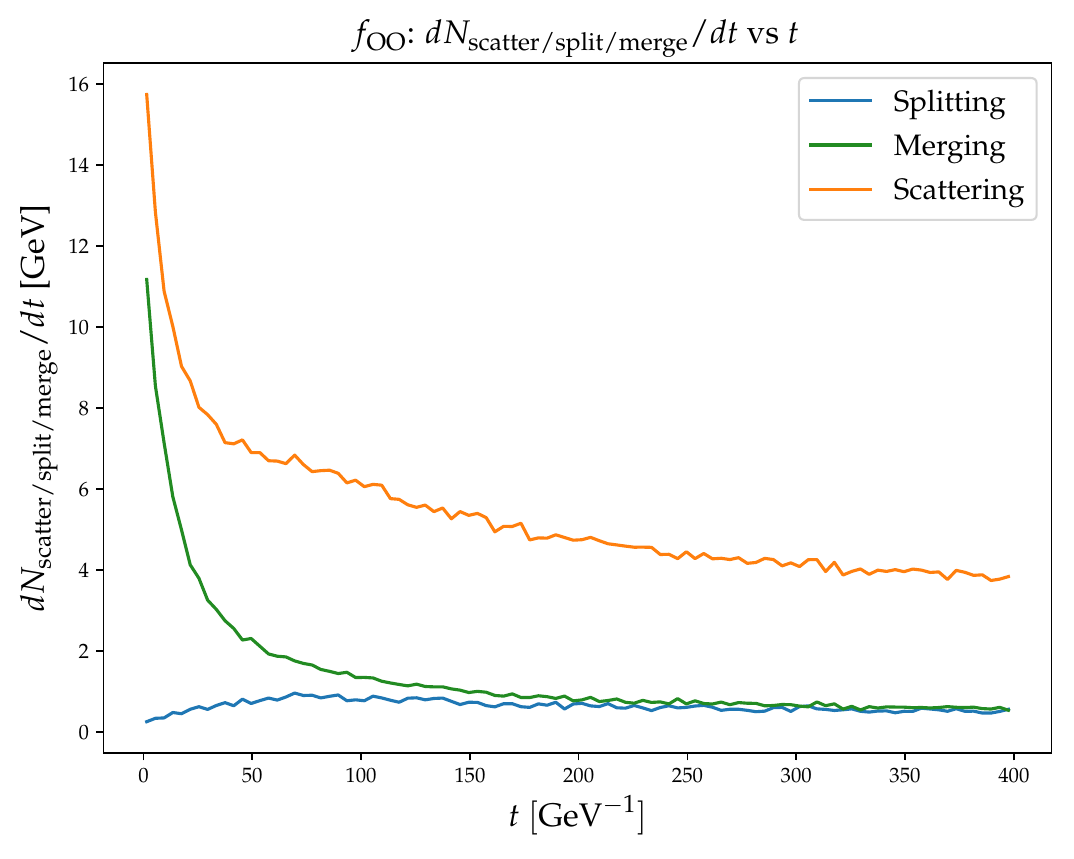}
    \caption{Scattering, splitting and merging rates as functions of time, for the overoccupied system in \textsc{Alpaca}.}
    \label{fig:thermalOO_rates}
\end{figure}

Lastly for the overoccupied case, we see the scattering, splitting and merging rates in \Fig{fig:thermalOO_rates} for the same initial setup in \textsc{Alpaca}. As expected, the system starts with much larger rate of mergings compared to rate of splittings, but eventually stabilizes at the same rates. The number of gluons in the system decreases very quickly in the first $\unit[70]{GeV^{-1}}$, followed by a longer phase of kinetic equilibration driven by elastic scattering.

\subsection{Isotropization}

As a first study of isotropization in \textsc{Alpaca}, we use the Color Glass Condensate-like initial conditions given in \cite{Kurkela:2015qoa, Lappi:2011ju}, 

\begin{equation}
    f_{\mathrm{CGCl}} = \frac{2A}{\lambda}\frac{1}{p_\xi}e^{-2p_\xi^2/3}
\end{equation}
where $A$ is controls the occupancy of the system and $\xi$ the longitudinal momentum asymmetry. The rescaled momentum is defined as $p_\xi = \sqrt{p_\perp^2 + \xi^2p_z^2}/\langle p_\perp \rangle$ where $\langle p_\perp \rangle$ is the initial average transverse momentum.

We use this form as an anisotropic initial condition (again for a purely gluonic system) to see how the system isotropizes. We here disregard longitudinal expansion and treat the system as spatially homogeneous.
This is, however, a system for which there are no analytical answers known for e.g. equilibration times. We instead treat this as a first look at the qualitative behaviour of the istropization process in \textsc{Alpaca}. We will focus on the initial parameters $\lambda=1$, $A=0.5$, $\xi=4$, $\langle p_\perp \rangle = 1.8$ \GeV. The events are initialized in a box of size $67$ $\GeV^{-3}$ with a momentum cutoff $p_{\mathrm{min}}=0.075$. 

\begin{figure}[tbp]
    \centering
    \includegraphics[width=1\linewidth]{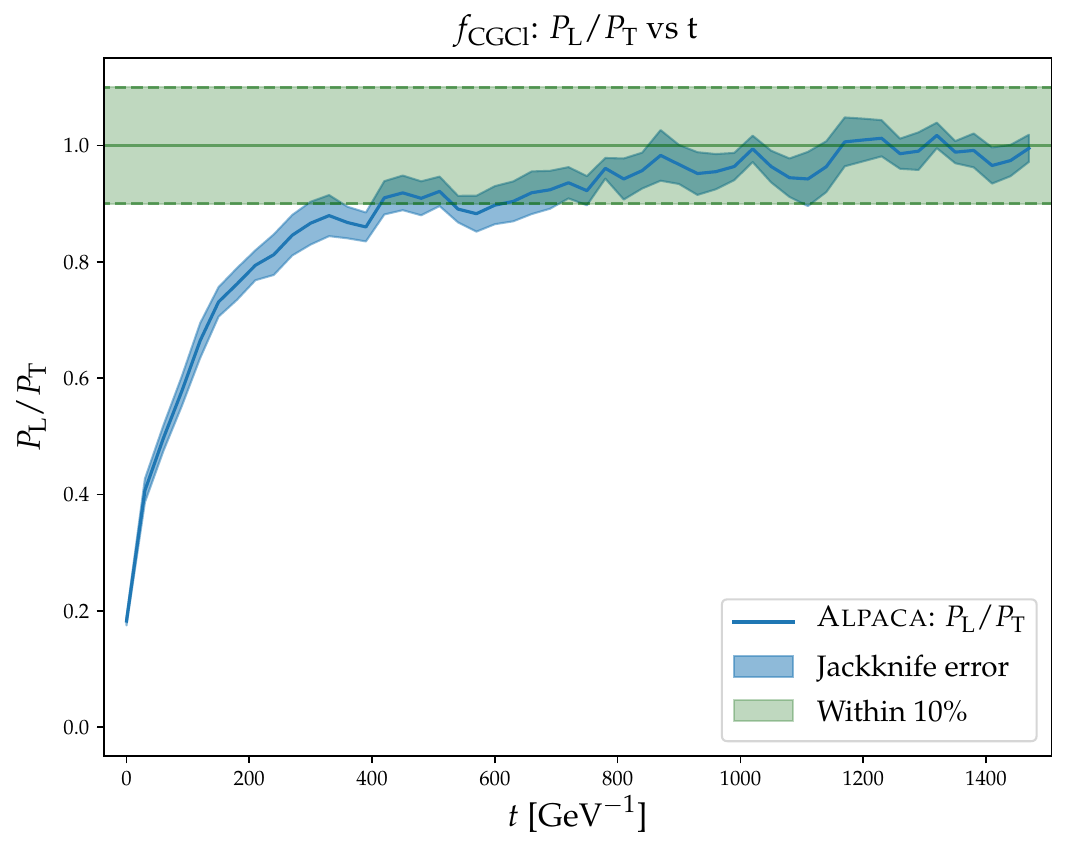}
    \caption{The ratio of longitudinal and transverse pressure, $P_{\mathrm{L}}/P_{\mathrm{T}}$, shown as a function of time. The blue band corresponds to the error of splitting the mean ratio of each event into ten subsamples and then taking using the Jackknife method to extract the error of the subsamples. The green ban corresponds to values within $10\%$ of unity.}
    \label{fig:anisotropic_PLTPT}
\end{figure}

\begin{figure*}[tbp]
    \centering
    \includegraphics[width=0.9\linewidth]{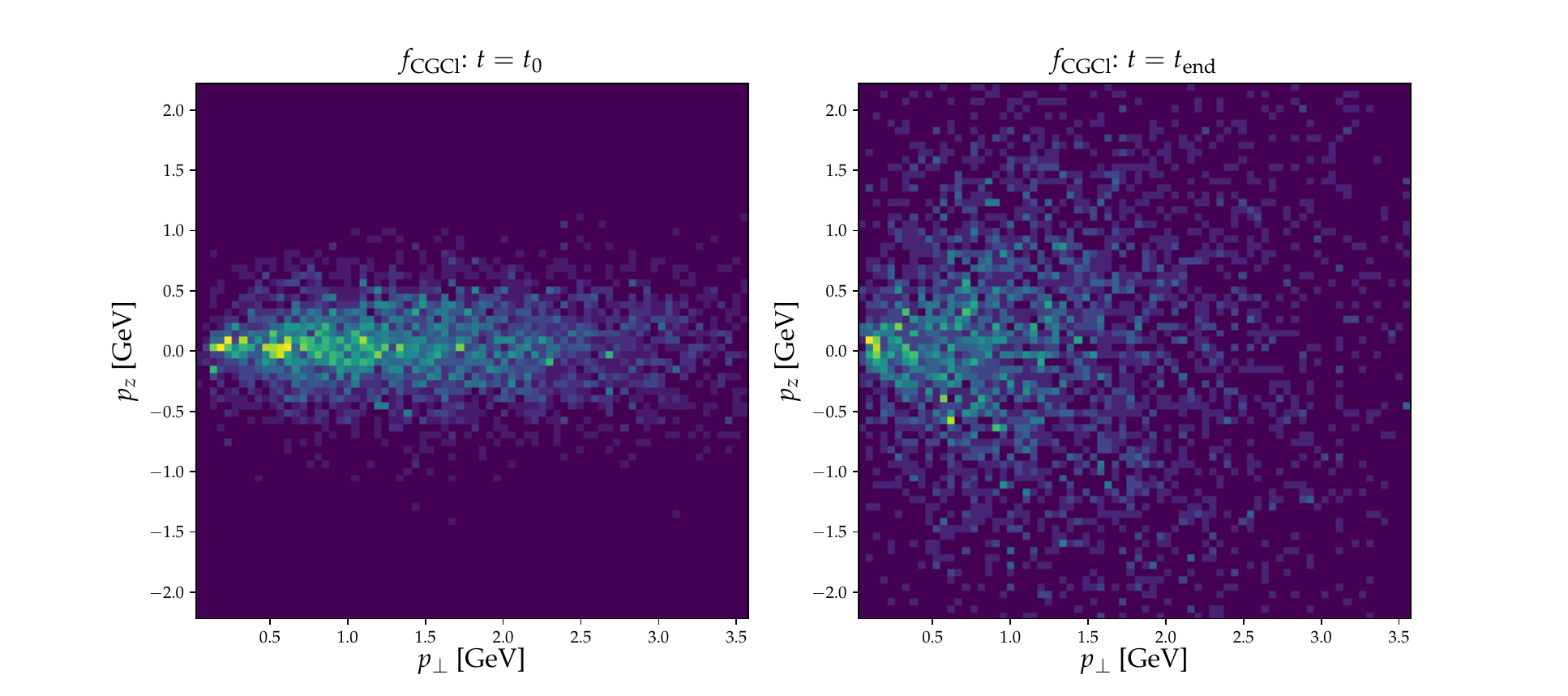}
    \caption{Heatmap of the initial and final $p_\perp$ and $p_z$ for the Color Glass Condensate-like initial conditions in \textsc{Alpaca}.}
    \label{fig:anisotropic_pTpz}
\end{figure*}

In \Fig{fig:anisotropic_PLTPT} the ratio of longitudinal and transverse pressure, defined as

\begin{equation}
    \frac{P_{\mathrm{L}}}{P_{\mathrm{T}}} = \frac{\int\frac{\d^3 \p}{(2\pi)^3 |\p|}p_z^2f(t,\p,\x)}{\int\frac{\d^3 \p}{2(2\pi)^3 |\p|}(p_x^2+p_y^2)f(t,\p,\x)},
\end{equation}
is shown. We see that the system indeed isotropizes, as expected. The process moves with comparatively high rate up to within $10\%$ of completely isotropic, at around $400$ $\GeV^{-1}$, and then slows down considerable for the last part of the evolution to a ratio of unity. A heatmap of the distribution of longitudinal and transverse momenta, at the beginning and end of the run, can be seen in \Fig{fig:anisotropic_pTpz}.

\section{Conclusions and outlook}
\label{section:conclusions}

In this work we have presented further validation of the newly introduced AMY kinetic theory parton cascade \textsc{Alpaca}, through looking at thermalization in overoccupied systems and isotropization in homogeneous anisotropic systems.

The evolution of the initial overoccupied distribution in \textsc{Alpaca} is shown to agree well with known analytical results for the same initial distribution. We see the thermalization happen with the correct equilibration time. We also see the correct qualitative behaviour of the evolution of the weighted distribution $(p/T)^3f(p)$. Lastly, we also observe the expected behaviour of the splitting, merging and scattering rates, with a much larger merging than splitting rate initially, followed by a phase of slower kinetic equilibration driven mostly by elastic scattering.

For the homogeneous anisotropic initial conditions sampled in a finite box with periodic boundary conditions, we do not have any analytical results to compare to. Instead, we treat this as a first step of validating the isotropization of \textsc{Alpaca}, and instead study the general behaviour of the longitudinal and transverse pressure in this setup. We find that the system isotropizes as expected in terms of the ratio $P_{\mathrm{L}}/P_{T}$ reaching unity, with the evolution to reach a ratio of $0.9$ being almost twice as fast as that of going from a ratio of $0.9$ to $1$.

The natural next step for the studies of equilibration processes in \textsc{Alpaca} is to look at thermalization of underoccupied systems. This is a well studied process and there are analytical results to compare the results of \textsc{Alpaca} to, e.g \cite{Kurkela:2014tea, Fu:2021jhl}.
An extended study of isotropization is also needed, where it would be of interest to reproduce the initial setup of \cite{Kurkela:2015qoa} by including longitudinal expansion and compare to the corresponding analytical results found in that paper.

\section*{Acknowledgments}
The authors would like to thank Aleksi Kurkela for helpful discussions and explanations of their calcuation.
This study is part of a project that has received funding from the European Research Council (ERC) under the European Union's Horizon 2020 research and innovation programme  (Grant agreement No. 803183, collectiveQCD).

\bibliography{references}

\appendix

\end{document}